\documentclass[journal]{IEEEtran}
\usepackage{graphicx}
\usepackage{amsmath}
\usepackage{dsfont}
\usepackage[english]{babel}
\usepackage{amsthm}
\usepackage[dvipsnames]{xcolor}
\usepackage{amssymb}
\usepackage{xfrac}
\usepackage{units}
\usepackage{booktabs}
\usepackage{balance}
\usepackage{url}

\theoremstyle{definition}
\newtheorem*{definition*}{Definition}

\newcommand{\MI}{\ensuremath{\mathcal{I}}}
\newcommand{\E}{\ensuremath{\mathcal{H}}}

\begin{document}

\title{Adaptive Transform Coding for Semantic Compression}

\author{Andriy Enttsel, \IEEEmembership{Member, IEEE}, Vincent Corlay, \IEEEmembership{Member, IEEE}
\thanks{A. Enttsel and V. Corlay are with Mitsubishi Electric R\&D Centre Europe, Rennes, 35700, France (e-mail: a.enttsel@fr.merce.mee.com, v.corlay@fr.merce.mee.com).}
}

\maketitle

\begin{abstract}
Visual data compression is shifting from human-centered reconstruction to machine-oriented representation coding. In this setting, an image is often mapped to a compact semantic embedding, which is then compressed and transmitted for downstream inference. We propose an adaptive transform-coding method for semantic-feature compression motivated by the conditional rate–distortion function of a Gaussian mixture model. The scheme uses mode-dependent transforms and quantizers selected according to the inferred source component, enabling more efficient coding of heterogeneous feature distributions. Evaluations on features from widely used vision backbones and foundation models show that the proposed method outperforms or is competitive with state-of-the-art neural compression methods while preserving flexibility and interpretability.
\end{abstract}


\IEEEpeerreviewmaketitle

\section{Introduction}

Edge devices, mobile phones, and autonomous vehicles generate massive amounts of visual data. While such data have traditionally been compressed and transmitted to the cloud for storage and processing, vision models are increasingly being deployed at the edge, where they extract semantically meaningful representations locally. In many practical settings, these representations must still be transmitted to a server for aggregation, coordination, or fusion with information from other agents \cite{Hua2023ACM}. This shift makes compression of learned representations, rather than raw pixels, an important problem.

In contrast to conventional visual compression, which is designed to preserve perceptual fidelity, task-oriented source coding aims to preserve performance on downstream vision tasks such as classification \cite{Singh2020ICIP,Matsubara2022WACV},
and object detection \cite{Yuan2022MIPR}. Existing methods are often effective, but many are tied to a particular task, or to a fixed set of tasks, and may require retraining when the task or operating conditions change \cite{Zamir2018CVPR,Alvar2021TIP,Matsubara2025WACV,Guo2025ILCR}. This limitation motivates task-agnostic approaches to feature compression.

This direction is also reflected in current standardization efforts. JPEG AI has identified the preservation of useful information for machine analysis as an important long-term objective \cite{Ascenso2023Multim,Alshina2024Multim}, while MPEG Feature Coding for Machines (MPEG FCM)\footnote{\url{https://www.mpeg.org/standards/MPEG-AI/4/}} focuses on compressing neural-network features rather than raw pixels. 

At the same time, recent multimodal foundation models provide generic semantic embeddings that transfer across tasks and domains \cite{Radford2021PMLR,Liu2023NIPS}. Building on this idea, \cite{Shen2025MLSP} introduced a task-agnostic semantic compression framework based on CLIP \cite{Radford2021PMLR} and showed strong generalization across data distributions and downstream embedding-based tasks, such as classification and retrieval.

In this letter, we revisit semantic feature compression from a new perspective. Instead of proposing a learned neural codec, we demonstrate that classical transform-coding principles \cite{Goyal_2001SPM}, at the core of JPEG \cite{WallaceTCE1992, SkodrasSPM2001}, are even more effective for semantic embeddings, including CLIP representations and deep features from standard vision backbones. Our approach builds on adaptive transform coding \cite{Dony_1995TIP,Archer_2000NIPS,Archer_2004TSP} and is grounded in the conditional rate--distortion function of a Gaussian mixture model (GMM). By combining rate--distortion results for Gaussian sources \cite{Berger1971,Cover2006} with conditional rate--distortion theory \cite{Gray_1972TR}, we arrive at a mode-dependent coding scheme in which both the transform and scalar quantizers are selected based on the inferred mixture component.

The resulting method is simple, interpretable, and flexible.
Moreover, experiments on semantic embeddings extracted from a CLIP foundation model show that the proposed scheme outperforms not only conventional principal component analysis (PCA)-style baselines \cite{Jolliffe_2002} used in prior feature-coding work, but also neural network-based semantic codecs.

Our contributions are summarized as follows:
\begin{itemize}
\item We derive a genie-aided upper bound to the rate--distortion function of a GMM based on mode conditioning, governed by a single reverse-water-filling parameter.
\item We translate this analysis into a practical adaptive transform coder for semantic feature compression using mode-dependent Karhunen--Lo\`eve transforms (KLTs) \cite{Goyal_2001SPM} and scalar Lloyd--Max quantizers \cite{LloydTIT1982}.
\item We show that, for CLIP embeddings, this simple neural-network-free scheme outperforms or is competitive with learned semantic compression baselines.
\end{itemize}

\section{Theoretical Model}

Our scheme falls within the framework of \emph{conditional rate--distortion theory} \cite{Gray_1972TR}. In contrast to conditional entropy coding in classical and neural source coding \cite{Minnen2018NIPS,Ballé_2021JSTSP,Slepian1973TIT}, where side information serves only to improve entropy modeling, our scheme uses it also to drive lossy transform selection.

\begin{figure}
        \centering
        \includegraphics[width=\linewidth]{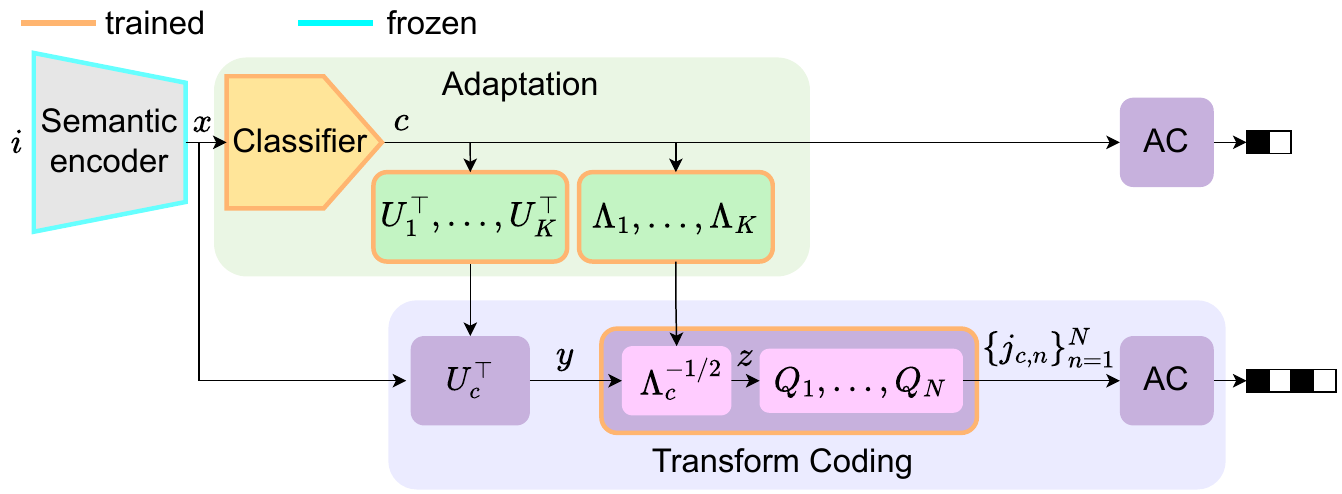}

	\caption{Illustration of the semantic compressor. The semantic encoder extracts the image semantic information. The adaptation layer selects the most suitable transform coding path, comprising orthogonal transformation, quantization, and arithmetic coding (AC). Offline, only the adaptation layer is fitted and is completely defined by the GMM.}
    \label{fig: scheme}
\end{figure}

\subsection{Conditional Rate--Distortion Function}
Consider an $N$-dimensional source $X$. Under the MSE distortion measure, its rate--distortion function is \cite[Chapter~10]{Cover2006}
\begin{equation}
\label{eq: rate-distortion}
R_X(D)
=
\inf_{p(\hat{X}|X)}
\MI \big(X;\hat{X} \big)
\quad \text{s.t.} \quad
\mathbb{E}\Big[ \big\|X-\hat{X}\big\|^2 \Big]\le D,
\end{equation}
where $\MI(X;\hat{X})$ is the rate, $\mathbb{E}[\|X-\hat{X}\|^2]$ is the mean squared error (MSE) distortion, and $p(\hat{X}|X)$ is the channel.

Let $C$ be a latent variable such that $C \rightarrow X \rightarrow \hat{X}$. By the chain rule for mutual information,
\begin{align}
\label{eq: equation a}
\MI(X;\hat{X})
&= \MI(X,C;\hat{X}) - \MI(C;\hat{X}\mid X) \\
\label{eq: equation b}
&= \MI(C;\hat{X}) + \MI(X;\hat{X}\mid C) - \MI(C;\hat{X}\mid X) \\
\label{eq: equation c}
&= \MI(C;\hat{X}) + \MI(X;\hat{X}\mid C),
\end{align}
where \eqref{eq: equation c} follows from the Markov condition, which implies $\MI(C;\hat{X}\mid X)=0$. Therefore \eqref{eq: rate-distortion} becomes
\begin{align}
R_X(D)
&=
\inf_{p(\hat{X}\mid X)}
\big[\MI(C;\hat{X}) + \MI(X;\hat{X}\mid C)\big] \\
\notag
&\text{s.t.} \quad \mathbb{E}\Big[ \big\|X-\hat{X}\big\|^2 \Big]\le D.
\end{align}
Under the assumption that $C$ is revealed to the encoder by a genie, this reformulation does not reduce the rate of the original problem, but suggests a coding scheme in which $C$ is first transmitted losslessly to the decoder and $X$ is then encoded conditionally on $C$.

Accordingly, the conditional rate--distortion function is \cite{Gray_1972TR}
\begin{equation}
\label{eq: conditional rate-distortion}
R_{X \mid C}(D)
=
\inf_{p(\hat{X}|X,C)}
\MI(X;\hat{X}\mid C)
\quad \text{s.t.} \quad
\mathbb{E}[\|X-\hat{X}\|^2]\le D.
\end{equation}
This yields the sandwich bound
\begin{equation}
\label{eq:upper bound}
R_{X \mid C}(D) \le R_X(D) \le R_{X \mid C}(D) + \E(C),
\end{equation}
where the genie-aided upper bound corresponds to lossless transmission of $C$ followed by conditional coding of $X$ given $C$.

If $C \in \{1,\dots,K\}$, then \eqref{eq: conditional rate-distortion} decomposes as \cite[Theorem~5]{Gray_1972TR}
\begin{equation}
\label{eq: mixture rate-distortion}
R_{X \mid C}(D)
=
\inf_{\mathcal{D}}
\sum_{c=1}^K \pi_c R_{X \mid C=c}(D_c)
\quad \text{s.t.} \quad
\sum_{c=1}^K \pi_c D_c \le D,
\end{equation}
with $\mathcal{D}=\{D_1,\dots,D_K\}$. Thus, the problem reduces to allocating distortion across components. At the optimum,
\begin{equation}
\label{eq: weighted rate distortion}
R_{X \mid C}(D)
=
\sum_{c=1}^K \pi_c R_{X \mid C=c}(D_c),
\end{equation}
where the distortions satisfy the distortion budget $D=\sum_{c=1}^K \pi_c D_c$ and the equal slope conditions
\begin{equation}
\label{eq: equal slope}
R'_{X \mid C=1}(D_1)=\cdots=R'_{X \mid C=K}(D_K).
\end{equation}

\begin{figure*}
    \centering
    \includegraphics[width=\textwidth]{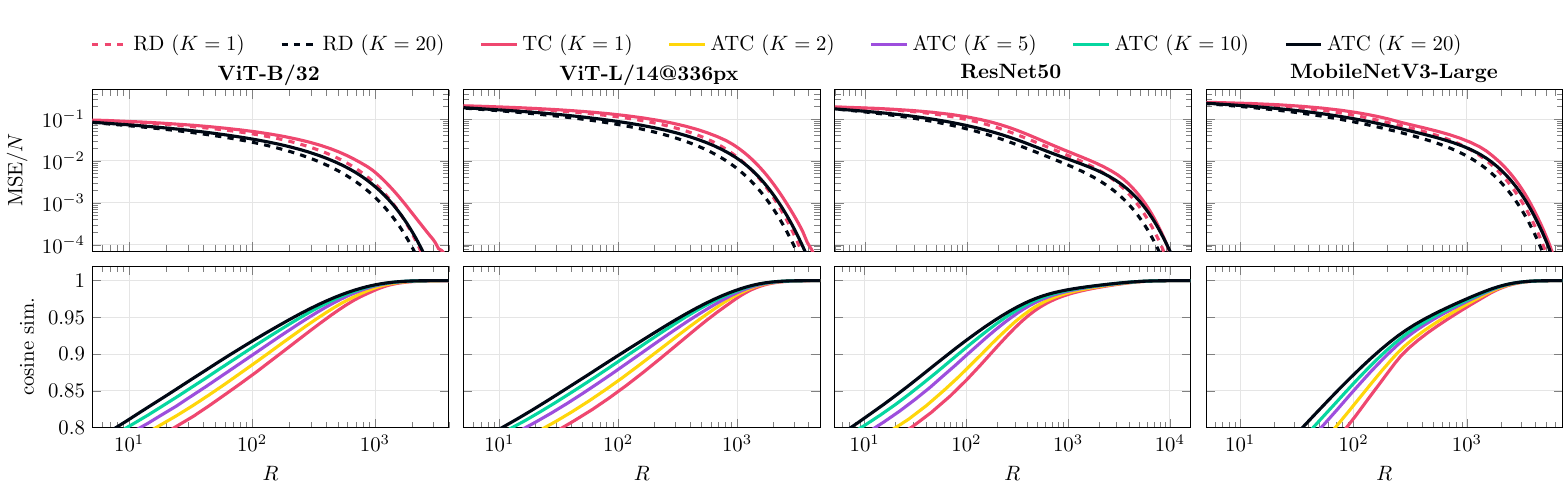}
    \caption{Rate--distortion (RD) performance in terms of normalized MSE (top) and cosine similarity (bottom) of the adaptive transform coding (ATC) scheme compared with the rate--distortion (RD) function upper bound and the non-adaptive transform-coding (TC) baseline for different feature extraction models.}
    \label{fig: rd}
\end{figure*}

\subsection{Gaussian Mixture Model}
Assume now that $X$ follows a $K$-component GMM 
\begin{equation}
    X \sim \sum_{c=1}^K \pi_c \mathcal{N}(\mu_c,\Sigma_c)
\end{equation}
with eigendecomposition $\Sigma_c = U_c \Lambda_c U_c^\top$ and $\Lambda_c=\operatorname{diag}(\lambda_{c,1},\dots,\lambda_{c,N})$. Conditioned on component $C=c$, the source is Gaussian, and its rate--distortion function is \cite[Theorem~10.3.3]{Cover2006}
\begin{equation}
R_{X \mid C=c}(D_c)
=
\frac{1}{2}\sum_{n=1}^N \log_2 \frac{\lambda_{c,n}}{D_{c,n}},
\end{equation}
where $D_{c,n} = \min\{\lambda_{c,n}, \theta_c\}$ and $\theta_c$ is the reverse water-filling parameter chosen such that $D_c = \sum_{n=1}^N  D_{c,n}$.
Moreover, it is straightforward to show that $R'_c(D_c) = -\nicefrac{1}{2\theta_c \ln{2}} $.
Hence, by the equal-slope condition in \eqref{eq: equal slope}, all active dimensions share the same reverse-water-filling parameter $ \theta_1=\cdots=\theta_K=\theta$.

Therefore, \eqref{eq: mixture rate-distortion} becomes
\begin{equation}
\label{eq: gaussian mixture rate-distortion}
R_{X \mid C}(D)
=
\frac{1}{2}\sum_{c=1}^K \pi_c \sum_{n=1}^N
\log_2 \frac{\lambda_{c,n}}{\min\{\lambda_{c,n},\theta\}},
\end{equation}
where $\theta$ controlling both inter- and intra-component distortion allocation is chosen such that 
 \begin{equation}
     D=\sum_{c=1}^K \pi_c \sum_{n=1}^N \min\{\lambda_{c,n},\theta\}.
 \end{equation}

Finally, if $C$ is transmitted losslessly, the total rate of the factorized scheme is given by the right-hand side of \eqref{eq:upper bound}
\begin{equation}
\label{eq: approx}
\tilde R_X(D)
=
R_{X \mid C}(D) - \sum_{c=1}^K \pi_c \log_2 \pi_c.
\end{equation}

\section{Semantic Adaptive Transform Coding}

We now describe the practical coding scheme motivated by the analysis above.

\subsection{Offline design stage}

Given an image $I$, we first extract the embedding $X=f(I)$ using a frozen semantic encoder.  
We then fit\footnote{In the main approach, the GMM is fitted in an unsupervised manner. Possible supervised variants are discussed in the Appendix.} a GMM with $K$ components to the embeddings obtained on a dataset representative of $I$.

For each component $c$, after eigendecomposition of the covariance matrix $\Sigma_c$, the transform $U_c^\top$ is the component-specific KLT \cite{Goyal_2001SPM}.  After whitening, the corresponding coefficients become 
\begin{equation}
    Z_c = \Lambda_c^{-1/2} U_c^\top (X-\mu_c).
\end{equation}
For coefficient $Z_{c,n}$, given a global quality parameter $\theta$, the target distortion in the whitened domain is set by the reverse-water-filling rule $d_{c,n} = \min\{1,  \theta / \lambda_{c,n}\}$.

To implement this allocation, we use scalar Lloyd--Max quantizers \cite{LloydTIT1982}. First, we define a set of admissible numbers of quantization levels $\mathcal{L}=\{L^{(i)}\}_{i=1}^B$ with the corresponding MSE distortion
\begin{equation}
d_Q(L) = \mathbb{E}\left[(Z-Q_L(Z))^2\right],
\qquad L \in \mathcal{L}
\end{equation}
where $Q_{L}$ denotes the $L$-level Lloyd--Max quantizer
for a standard Gaussian source $Z\sim\mathcal{N}(0,1)$. Then, for each $(c,n)$, we choose the coarsest admissible quantizer as
\begin{equation}
L_{c,n}^\star
=
\min\left\{
L \in \mathcal{L} \;:\;  d_Q\left( L \right) \le d_{c,n}
\right\}.
\label{eq:quantizer-selection}
\end{equation}

\subsection{Online coding stage}

At encoding time, given an image realization $i$, we compute its embedding $x=f(i)$ and 
then for each cluster $c$ we compute the whitened vector 
\begin{equation}
    z_c = \Lambda_c^{-1/2} U_c^\top (x-\mu_c).
\end{equation}

Since the true component is unknown, we estimate it as the maximum-a-posteriori GMM component, or equivalently:
\begin{equation}
    \hat{c} = \arg\min_c \left\{ \| z_c \|_2^2 + \sum_{n=1}^N \log \lambda_{c, n} - 2\log \pi_c \right\}.
\end{equation}

Conditioned on $\hat c$, each scalar coefficient is quantized using the preselected Lloyd--Max quantizer with $L_{\hat c,n}^\star$ levels:
\begin{equation}
\hat{z}_{\hat c,n} = q_{j_{\hat c,n}} = Q_{L_{\hat c,n}^\star}\left( z_{\hat c,n} \right),
\end{equation}
where $q_{j_{\hat c,n}}$ is one of the $L_{\hat c,n}^\star$ reproduction centroids. Under the Gaussian model, the probability of output index $j$ is
\begin{equation}
\mathbb{P}\left(J_{\hat c,n}=j \mid C=\hat c;L_{\hat c,n}^\star \right)
=
\Phi \left(b_{j+1})-\Phi(b_j \right),
\end{equation}
where $\Phi(\cdot)$ is the standard Gaussian cumulative distribution function and $\{b_j \mid j=1, \dots, {L_{\hat c,n}^\star+1} \}$ are the quantizer decision boundaries, with $b_1=-\infty$ and $b_{L_{\hat c,n}^\star+1}=+\infty$. These probabilities are used for arithmetic coding of the quantizer indices \cite{Rissanen1976IBMJRD,Witten1987ACM}.

The mode index $\hat c$ is encoded separately using an arithmetic code based on the mixture weight $\pi_{\hat c}$. At the decoder, $\hat c$ and $\hat z_{\hat c}$ are recovered, and the feature vector is reconstructed as 
\begin{equation}
    \hat x = U_{\hat c} \Lambda_{\hat c} ^{1/2} \hat z_{\hat c}  + \mu_{\hat c}. 
\end{equation}
The overall scheme is summarized in Fig.~\ref{fig: scheme}.

\subsection{Discussion}
In classical image compression, transform coding is typically applied to small, assumed to be Gaussian patches, and adaptive extensions often provide limited gains relative to their complexity \cite{Archer_2004TSP}. Larger gains can be achieved by nonlinear transform coding schemes \cite{Ballé_2021JSTSP}, such as those underlying JPEG AI \cite{Ascenso2023Multim,Alshina2024Multim}. They rely on a side-information signal, called the \textit{hyperprior}, conditioned on which the residual signal can be modeled by a factorized Gaussian distribution \cite{Minnen2018NIPS}.

Semantic embeddings are better suited to a mixture-based treatment. Unlike full images in the pixel domain, they are lower-dimensional and more structured, which makes a Gaussian-mixture model more plausible \cite{Lee2018NIPS}. In this sense, the inferred component in our scheme plays a role analogous to that of the hyperprior.


The dominant encoder-side computational and storage costs of the scheme scale as $O(N^2K)$. These costs can be reduced to $O(NM + M^2K)$ by first applying a global PCA-based projection that maps $X$ from $N$ to $M$ dimensions, and then fitting the GMM in the reduced $M$-dimensional space. A complete description and analysis of this variant are provided in the Appendix.


Finally, the method readily adapts to different rate--distortion trade-offs, since changing the quality level only requires updating the quantization map.

\begin{figure*}
    \centering
    \includegraphics[width=\textwidth]{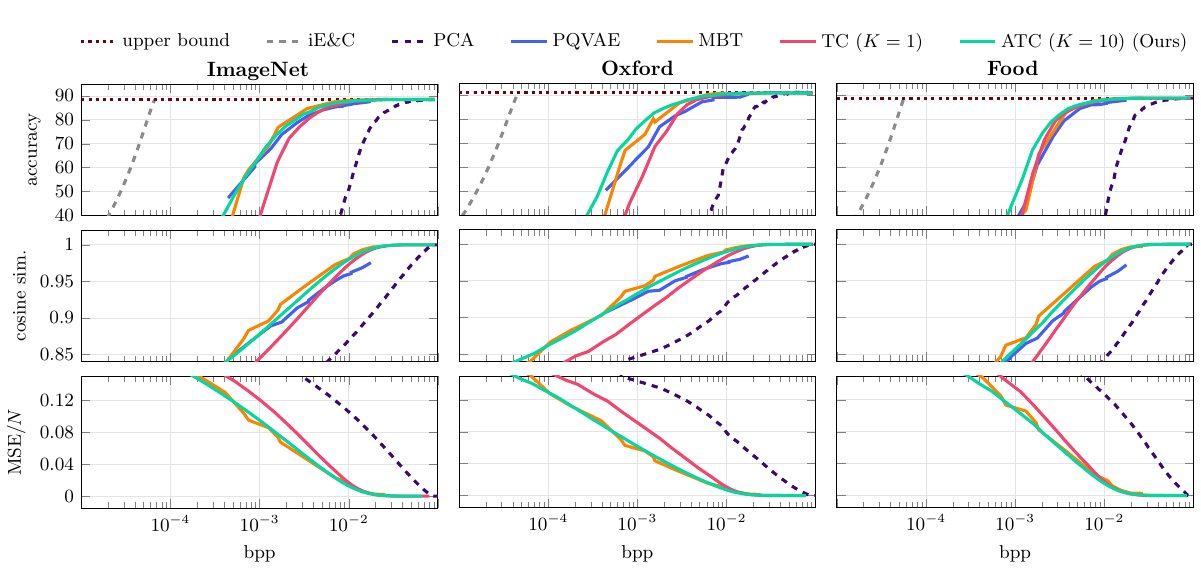}
    \caption{Rate in bits per pixel versus zero-shot classification accuracy (top) and cosine similarity (bottom) on three datasets for the proposed adaptive scheme (ATC), PQVAE, non-adaptive transform-coding (TC) baseline, PCA-based compression and the indirect estimate-and-compress bound (iE\&C).
    }
    \label{fig: soa}
\end{figure*}

\section{Numerical Results}

\subsection{Rate--distortion performance}

We first evaluate the proposed scheme in terms of the rate--distortion trade-off. The rate is measured as
\begin{equation}
\label{eq: empirical rate}
R = \sum_{n=1}^N \mathbb{E}\left[-\log_2 \mathbb{P}\left(J_{\hat c,n}=j \mid C=\hat c \right)\right]
- \mathbb{E} \left[ \log_2 \pi_{\hat c} \right],
\end{equation}
where the first term is the expected code length of the entropy-coded quantizer indices, and the second term is the expected code length of the inferred mixture-component index. 

We consider four semantic encoders: two vision backbones, ResNet50 and MobileNetV3-Large, and two CLIP models, ViT-B/32 and ViT-L/14@336px, and extract embeddings of dimension $N \in \{2048,960,768,512\}$, respectively, from 200,000 ImageNet training images. For the vision models, the embeddings are taken from the average-pooling layer before the classification head, whereas for the CLIP models, we use the non-normalized encoder outputs. We then fit GMMs with $K \in \{1,2,5,10,20\}$ components to these embeddings\footnote{To ensure positive-definite covariance estimates, we add diagonal regularization. The regularization weight is $10^{-6}$, except for the cases ViT-L/14@336px with $K \in \{10, 20\}$, for which it is $10^{-5}$.}.

The results are shown in the top plot of Fig.~\ref{fig: rd}, where the distortion is measured by the normalized MSE.
The black solid curves correspond to the adaptive scheme with $K=10$, while the red solid curves denote the non-adaptive baseline, i.e. $K=1$, for different values of the quality parameter $\theta$. 
As a lower bound for the practical scheme, we use the rate--distortion function in \eqref{eq: approx} estimated on the training set embeddings and represented with dashed curves.

The first observation is that adaptation consistently improves rate--distortion performance across all encoders. Second, a visible gap remains between the empirical curves and the theoretical bound, especially in the medium-rate regime. This is expected, since the rate--distortion function is asymptotic, whereas the proposed scheme relies on one-shot scalar quantization. Moreover, the fitted GMM remains only an approximate model of the embedding distribution.

To further assess semantic preservation and the effect of the number of mixture components, the bottom row of Fig.~\ref{fig: rd} reports the trade-off in terms of cosine similarity, a standard metric for comparing semantic embeddings, for different values of $K$. Here, the benefit of adaptation is even more pronounced: cosine similarity improves monotonically with $K$, yielding gains of up to 5\% for $K=20$ relative to $K=1$.


\subsection{Comparison with the state-of-the-art}

We compare our scheme with PQVAE, the recent neural feature-compression method of~\cite{Shen2025MLSP}, and with MBT, a custom adaptation of the neural mean-and-scale hyperprior codec introduced in~\cite{Minnen2018NIPS}. Since the original MBT codec was designed for image compression, we adapt it to semantic feature compression, as detailed in the Appendix. For direct comparability with~\cite{Shen2025MLSP}, we follow their experimental protocol: the compressor is trained on ViT-L/14@336px embeddings extracted from 40,000 ImageNet images, obtained by randomly selecting 800 classes.

We first evaluate the trade-off between compression rate, measured in bits per pixel (bpp)\footnote{$\mathrm{bpp} = \text{average bit rate per image} / \text{\# pixels per image}$. Since entropy coding is not explicitly implemented in \cite{Shen2025MLSP}, we use a fixed-length code (FLC) for all codecs except MBT in this experiment, for which the rate is measured similarly to \eqref{eq: empirical rate}.}, and downstream zero-shot classification accuracy is evaluated on the remaining validation images.
To assess out-of-distribution generalization, we also report results on Oxford-IIIT Pet \cite{Parkhi2012CVPR} and Food-101 \cite{BossardECCV2014}.

The results are shown in the top plot of Fig.~\ref{fig: soa}. The green curve corresponds to the proposed adaptive scheme. On the considered ImageNet subset, the accuracy stays close to the uncompressed upper bound of $88.5\%$ (dotted line) at high bit rates and degrades gradually as the rate decreases. This behavior is similar to that of MBT. In the medium-bpp regime, our method consistently outperforms PQVAE (blue). The non-adaptive transform-coding baseline (red) suffers a larger performance drop, but still reduces the bitrate by roughly a factor of eight compared with PCA, whose components are represented with $16$-bit precision (light blue).

On Oxford-IIIT Pet, our scheme shows better generalization than both PQVAE and MBT; in the high-bpp regime, even the non-adaptive baseline outperforms PQVAE. On Food-101, the adaptive scheme again surpasses the neural schemes, while the non-adaptive baseline is comparable or superior depending on the operating point.

The gray curve shows the theoretical model-aware rate--accuracy bound of the idealized \textit{indirect estimate and compress} (iE\&C) scheme \cite{Enttsel2026arxiv}, which first estimates the class-relevant information under the assumed CLIP model and then compresses it using an information-theoretically optimal task-specific codec. Because our method and the competing baselines are task-agnostic, their gap to the iE\&C bound represents the rate cost of task agnosticism.

The second row of Fig.~\ref{fig: soa} shows the trade-off in terms of cosine similarity. Since PQVAE is trained with a cosine-similarity objective, this comparison is particularly relevant. The same overall trends are observed: the proposed adaptive scheme remains competitive with, and often outperforms, PQVAE across operating points. Although the classical approaches are optimized for MSE rather than cosine similarity, they are still able to achieve high cosine similarity scores in the high-rate regime, in contrast to PQVAE. 
MBT attains comparable or better cosine-similarity performance than our adaptive scheme. This advantage is partly due to the use of entropy coding for MBT, whereas the transform-coding schemes are evaluated with fixed-length coding in this comparison. Indeed, when entropy coding is also applied to the adaptive scheme, as shown in the last row of the figure, its normalized-MSE performance substantially overlaps with that of MBT.

\section{Conclusion}
We proposed a task-agnostic semantic feature compression method based on adaptive transform coding, guided by a rate--distortion analysis of Gaussian mixture models. By selecting transforms and quantizers according to the inferred feature mode, the method outperforms both non-adaptive coding and a state-of-the-art semantic codec in preserving semantically relevant information, offering a simple, interpretable, and effective non-neural alternative to standard PCA-based solutions.

\section*{Appendix}
\subsection{Supervised Gaussian Mixture Model}

The mixture components can be defined either in an unsupervised or in a supervised manner. In the unsupervised case, the component index is obtained directly by GMM clustering in the feature space. In the supervised case, the component index can be derived from available semantic information, with the component statistics defined as follows
\begin{align}
\label{eq: gmm_supervised}
    \Sigma_c &= \mathbb{E}\bigl[(X_c-\mu_c)(X_c-\mu_c)^\top\bigr] = U_c \Lambda_c U_c^\top, \\
    \mu_c &= \mathbb{E}[X_c], \qquad
    \pi_c = \mathbb{P}(C=c).
\end{align}

For ImageNet, the semantics can be defined through a predefined grouping of fine-grained classes into a smaller number of coarse semantic categories, i.e., \emph{superclasses}. To estimate the quantities in \eqref{eq: gmm_supervised}, it is sufficient to define the component index as $C=s(\bar C)$, where $\bar C$ is the fine-grained label associated with $X$ and $s(\cdot)$ denotes the mapping from a fine-grained class to its corresponding superclass.

For a standard vision backbone trained for classification, let $h(f(i))$ denote the vector of logits for image $i$, where $f(\cdot)$ is the feature extractor and $h(\cdot)$ is the classification head. At the online coding stage, the component index is given by
\begin{equation}
    \hat c = s\!\left(\arg\max_{k} \, h_k(f(i))\right).
\end{equation}

For CLIP-based models, let $\{t_1,\dots,t_K\}$ be a set of candidate text prompts generated from labels $\{c_1,\dots,c_K\}$, and let $g(t_c)$ denote the corresponding precomputed text embedding. In the online phase, the component index is defined as the prompt whose text embedding is most similar to the image embedding $f(i)$, namely
\begin{equation}
    \hat c = \arg\max_{k \in \{1,\dots,K\}}
    \frac{f(i)^\top g(t_k)}{\|f(i)\|_2\,\|g(t_k)\|_2}.
\end{equation}
That is, $\hat c$ is the index of the text prompt with maximum cosine similarity to the image embedding.

\subsection{Complexity-aware scheme}
\label{subsec: comp scheme}
The definition of $\Sigma_c$ is quite general. In the main approach, it is obtained by fitting a GMM directly to the feature vector $X \in \mathbb{R}^N$. A more complexity-aware alternative is to first perform a global PCA on $X$. Let
\begin{equation}
\Sigma = \mathbb{E}\bigl[(X-\mu)(X-\mu)^\top\bigr] = U \Lambda U^\top,
\;\;\;
\mu = \mathbb{E}[X].
\end{equation}
Retaining the first $M$ principal directions, define
\begin{equation}
\label{eq: pca}
V = [u_1,\dots,u_M] \in \mathbb{R}^{N\times M},
\;\;
\bar X = V^\top (X-\mu) \in \mathbb{R}^M.
\end{equation}

A GMM is then fitted to $\bar X$. For component $c$, let $\bar \Sigma_c = \bar U_c \bar \Lambda_c \bar U_c^\top$
be the eigendecomposition of the component covariance in the reduced space. If $\bar \mu_c \in \mathbb{R}^M$ is the mean of component $c$ in the reduced space, the corresponding per-mode whitening transform is $Z_c = \bar \Lambda_c^{-1/2} \bar U_c^\top (\bar X-\bar \mu_c)$.
Substituting $\bar X $ defined in \eqref{eq: pca} gives
\begin{equation}
Z_c=
\bar \Lambda_c^{-1/2} \bar U_c^\top \bigl[V^\top(X-\mu)-\bar \mu_c\bigr]
=
\bar \Lambda_c^{-1/2} \tilde U_c^\top (X-\tilde \mu_c),
\end{equation}
with $\tilde \mu_c = \mu + V\bar \mu_c \in \mathbb{R}^N$, $\tilde U_c = V \bar U_c \in \mathbb{R}^{N \times M}$,
and where we used the fact that $V^\top V = I_M$.

This has the same whitening structure as in the original formulation. However, its implementation requires storing one global matrix $V \in \mathbb{R}^{N\times M}$, the global mean vector $\mu \in \mathbb{R}^{N}$, and, for each component, the matrix $\bar \Lambda_c^{-1/2} \bar U_c^\top \in \mathbb{R}^{M \times M}$ together with the vector $\bar \Lambda_c^{-1/2} \bar U_c^\top \bar \mu_c \in \mathbb{R}^{M}$.

To select $M$, we use the explained-variance criterion with threshold $\gamma$:
\begin{equation}
    M =
\min \left\{
m \in \{1,\dots,N\} :
\frac{\sum_{n=1}^{m}\lambda_n}{\sum_{n=1}^{N}\lambda_n} \ge \gamma
\right\}
\end{equation}

\begin{figure}
    \centering
    \includegraphics[width=\linewidth]{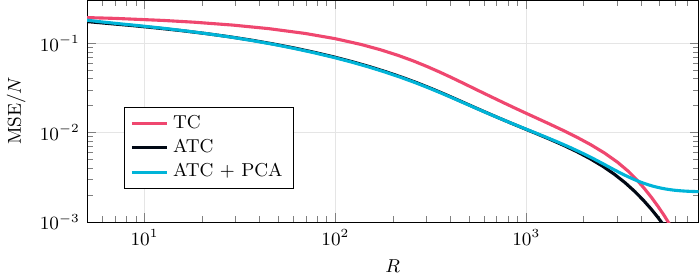}
    \caption{Rate--distortion performance in terms of normalized MSE of the adaptive transform coding (ATC) paired with PCA reduction technique. }
    \label{fig: ablation}
\end{figure}

In Fig.~\ref{fig: ablation}, we compare the main adaptive scheme with this reduced-complexity variant. We use ResNet50 ($N=2048$) as the semantic encoder, $K=20$, and otherwise follow the setup of Fig.~\ref{fig: rd}. 

The original scheme requires $N^2K + NK = 83.9$M parameters, while the PCA variant with $\gamma=0.99$ ($M=1330$) reduces this to $NM + N + (M+1)KM = 38.1$M parameters.

Across most of the rate--distortion curve, the reduced-complexity variant closely tracks the performance of the full adaptive scheme, with noticeable saturation only in the regime where conventional transform coding is already effective. 

\subsection{MBT implementation details}
The specialization of the MBT schemes is based on the implementation of {\tt mbt2018-mean} provided in the CompressAI PyTorch library \cite{Begaint_2020arxiv}. We replace the analysis and synthesis transforms, including those associated with the hyperprior, by architectures adapted to the compression of one-dimensional feature vectors.

\subsubsection{Architecture}
Given an input feature vector $x \in \mathbb{R}^{N}$, the main analysis transform $g_a$ maps $x$ to a latent representation $z \in \mathbb{R}^{M}$ through two linear layers with hidden dimension $H$ and an intermediate LeakyReLU activation, while the synthesis transform $g_s$ mirrors $g_a$  to reconstruct the feature vector from the quantized latent $\hat z$. Their architectures are
\begin{equation}
    g_a : \mathbb{R}^{N} \rightarrow \mathbb{R}^{H} \rightarrow \mathbb{R}^{M}, \qquad
    g_s : \mathbb{R}^{M} \rightarrow \mathbb{R}^{H} \rightarrow \mathbb{R}^{N}.
\end{equation}

To parameterize the entropy model, we use a hierarchical prior implemented by two additional two-layer perceptrons with an intermediate LeakyReLU activation. The hyper-analysis transform maps the latent vector $z$ to a lower-dimensional hyper-latent $w \in \mathbb{R}^{M/h}$, while the hyper-synthesis transform predicts the parameters of the conditional Gaussian model from the quantized hyper-latent $\hat w$:
\begin{equation}
    h_a : \mathbb{R}^{M} \rightarrow \mathbb{R}^{H/h} \rightarrow \mathbb{R}^{M/h}, \quad
    h_s : \mathbb{R}^{M/h} \rightarrow \mathbb{R}^{H/h} \rightarrow \mathbb{R}^{2M}.
\end{equation}
The output dimension $2M$ corresponds to the mean and scale parameters of the Gaussian conditional distribution associated with the $M$ latent coefficients.

\subsubsection{Optimization}
All models were trained for up to 1000 epochs using a batch size of 128. 
The main optimizer learning rate was initialized to $10^{-3}$ and reduced on plateau using a patience of 20 epochs and a reduction factor of 5. 
At most five learning-rate reductions were allowed during training. 
For the auxiliary parameters of the entropy model, the learning rate was fixed to $10^{-3}$. 
Early stopping was applied with a patience of 40 epochs.
The rate--distortion trade-off was controlled by the Lagrange multiplier weighting the MSE term, for which we considered the set $\{1,2,5\}\cdot 10^{\{0,1,2,3,4\}} \cup \{10^5,10^6\}$.
For the architecture parameters, we varied the hyperprior shrink factor as $h \in \{4,8,16\}$, the hidden dimension as $H \in \{N,2N\}$, and the bottleneck dimension as $M \in \{N/2, N\}$.
The results reported in Fig.~\ref{fig: soa} correspond to the models forming the MSE--rate Pareto frontier among all trained configurations.

\bibliographystyle{IEEEtran}
\bibliography{refs}

\end{document}